# What is the optimal anthropoid primate diet?

Hans Dehmelt

Department of Physics

University of Washington

Seattle, WA 98195

ABSTRACT     Following Socrates' advice "You should learn all you can from those who know. Everyone should watch himself throughout his life, and notice what sort of meat and drink and what form of exercise suit his constitution and he should regulate them in order to enjoy good health." Based on biological, chemical and physical considerations I have attempted to synthesize guide lines for an optimal diet from the vast literature. For an offshoot of the primate line it may be wise not to stray too far from the line's surprisingly uniform predominantly frugi- and herbi-vorous diet that is only lightly supplemented by hunted small mammals, eggs, nuts, insects etc. By dry weight raw wild fruit contains fats, proteins, carbohydrates, digested and undigested fiber in the approximate proportions 5 : 7 : 14 : 17 : 17. The fat component contains both *essential* fatty acids, about 23% linoleic and 16% alpha-linolenic, the latter severely lacking in Western diets. *The practical problem is how to as best as possible, but not religiously, approximate this diet with super-market items.*

GENERAL

The current Western diet clearly presents acute problems (1, 2, 3, 4, 5) that cry out for answers *now*. When I am dying from a faulty diet even educated guesses *now* would be much preferable to waiting, for example, for the results of double-blind diet studies to convince me that engaging in *any* activities to which evolution has not adapted me is risky business. Likewise, discovering previously overlooked physical, chemical and biological connections between food-related data appears of great value.

EVOLUTION & MODEL DIET

Adaptation by natural selection of random mutations is an imperfect very gradual process requiring very many generations to become effective. How far back in time must we go to find the ancient forbears that still were well adapted to their slowly changing environment? Changes in the basic body design of an established line of large mammals by environmental influences appear a near-impossibility. With the accelerating development of more and more tools and survival strategies there must have come a point when Homo x's changes of and to the environment outran evolution. It follows that at least since the beginnings of agriculture the populations of the industrialized countries can only have become more and more poorly adapted to their environment and are suffering the consequences. Our closest near-human relatives, chimpanzees, are ripe fruit specialists (5, 6); only when that is not available do they fall back on other diets. Because of that for simplicity we take the ripe fruit data in Table III of reference 7 (7, 8) as characterizing the anthropoid model diet. According to Fig.10 of reference 7 chimps feed four times as long on wild fruits and seeds as on leaves and pith combined and perhaps 5% of total feeding time on hunted small mammal prey, nuts, eggs, termites, etc. Averaged over annual variations by weight the principal ripe wild fruit component of their diet provides about 4.9 part fat (lipid), 9.5 parts crude protein (CP), 13.9 parts water soluble carbohydrates (WSC), 33.6 parts neutral detergent fiber (NDF). The chimps are able to digest the neutral detergent fiber (carbohydrate) in part (9, 10) : ~50% of NDF are transformed into short-chain fatty acids (SCFA) by hindgut fermentation and digested with an energy yield presumably similar to WSC. As suggested in references (11, 9) the CP value given above must be reduced to 6.7 parts and the (effective) WSC value must be increased to 30.7 parts. Accordingly, food energy would be derived from fat, protein, carbohydrate approximately in the ratio 11 : 6.7 : 30.7 = 455 cal : 280 cal : 1270 cal for a 2000 calorie diet from 51 g fat, 70 g protein and 320 g carbohydrate including digested fiber. Interestingly the 175 g intermediate digesta

SCFA, mostly acetic, proprionic and butyric acids (10), will have to be neutralized by the potassium and sodium of the food and form neutral soaps. Thus, presumably, the resulting soap solution daily will rinse out the hindgut and its circulatory system and assist in the emulsification of dietary fat. Surprisingly, this diet is very close in composition to that eaten by 3 small monkey species (7) living in the same African forest and wild howler monkeys (5) in Panama. The predominant fats (5) in the wild foods eaten by the Howler monkeys have been studied. They form a liquid mixture of oils and fats containing in various combinations 30% palmitic, 23% linoleic, 16% alpha linolenic, 15% oleic and 16% other fatty acids. All percentage values of food components listed are likely to be close to optimal for the primate body. Nevertheless, especially in man they exhibit great elasticity in the existence modes of Eskimos and starving human populations. The mathematical nature of optima suggests that they are uncritical. One might guess that over a range of a factor 2 smaller or larger deviations will not do any damage. Factors of 10 are another story. Choosing to go to chimpanzees for an appropriate model diet appears much less extreme in the light of the presumably much larger genetic difference between chimpanzees and monkeys and their nevertheless similar diets. On the contrary, humans should ask themselves how wise it is to deviate widely from a diet on which the primate line has flourished for many tens of millions of years and *produced them*.

THE FAT-WATER PROBLEMATIC

In the huge cell-state forming the human body the cells generate the energy for life by a low-temperature burn of foodstuffs fed to them by the blood circulating through nearby capillaries. To feed each individual cell the necessary oxygen and bite-sized food stuffs clearly creates terrifying supply and waste removal problems. The transport of fats or their components from the gut to the cells is made especially difficult as fat and water do not mix while amino acids and simple sugars do. The body solves the problem as well as it can in a similar way as fat stains in laundry are dissolved by soapy water. It emulsifies the liquid fat with the help of amphiphilic molecules that clad in a monolayer tiny < 0.1 mm-sized droplets of fat. These fragile packages (12) are then able to circulate in the blood through arteries, veins and, more or less, the capillaries. Thus, especially in regard to fats it appears wise not to deviate too much from the ancestral primate diet spelled out above. The respective melting points of the acids are: 63, -5, -11, 16 Centigrade. This demonstrates that the primate body can handle a fatty acid with a melting point considerably higher than its own normal temperature, at least when dissolved in the liquid ones and/or bound in oils/fats. Fatty acids with less than 16 or more than 18 carbon chains are uncommon (range 0 to 7%). Saturated and unsaturated fatty acids are almost equally balanced. The low melting points of the unsaturated fatty acids are associated with up to three kinks in their hydrocarbon chains that make it difficult for the weak intermolecular forces to form crystals and thus are presumably carried over to the fats containing them. The chimp body fat will presumably be a liquid mixture of fats and oils composed of various combinations of the above fatty acids in similar (13) overall proportions as in the diet. For a dietary fat mixture with a much higher melting point and of correspondingly higher viscosity on the other hand clogging up of arteries or capillaries might result. Mammary gland tissue is doubly challenged by the fat-water problematic as it has to produce the mother's milk. A recent study (14) shows that the average American gets 27 percent of his total daily energy from junk foods and an additional 4 percent from alcoholic beverages. The most conspicuous deficiency in his and all Western diets including that recommended by the "Food Pyramid" appears to be that of alpha-linolenic acid, one of the two fatty acids essential for life and one amply provided in the chimp diet (5). While rats can synthesize vitamin C the ancestors of primates lost this ability long ago. The vitamin is now "essential" for human life because in the past evolutionary history of primates there remained no need for such synthesis. Their arboreal environment provided it in abundance and, obviously, primates are very well adapted to this environment. In an analogous scenario presumably *all* animals very much earlier lost the ability to synthesize the two fatty acids essential for life when about 0.4 billion years ago the multicellular organisms branched into the passive dominant plant line and the mobile parasitic animal line of negligible biomass while the abundant vegetation continued to provided the animal line with ease. That in all this time no mutation was ever able to re-equip the animal body with the lost faculty underlines the utmost importance of these fatty acids to the primate body. Besides by their essential roles as cell constituents much of this importance may be explainable by low melting points and high fluidity of the oils built from them that "oil" the circulation by dissolving stoppages. By the same token the preponderance of low melting point poly-unsaturated fatty acids in cold water fish is unlikely to be an accident.

CORRECTION EXAMPLES

In a practical approach to an improved diet one might begin by identifying the largest, most damaging deviations from the model and then to *attempt to loosely* *approximate* the model diet with super-market items in the light of the above obvious great elasticity. To the author the sorest spots appear to be wrong dietary fat, much too low fiber and too much cane/beet sugar. Average persons having lived for decades according to the "Food Guide Pyramid", the misguided propaganda promoting hydrogenated fats or possibly on 30% junk food (14) might quickly bring the composition of their ~20 kg of body fat in line with that of the primate relatives. As in other mammals their depot fatty acid profile (13) is likely to closely resemble that of their diet and contain very little alpha-linolenic acid (ALA). From the above chimp data (5) one estimates a 16 % fat component of (ALA) as desirable. There is ~60 % ALA in flaxseed oil. Thus, quickly gaining the about 3 kg of the corresponding ALA based body fat component might be attempted – compare reference (15) -, by simultaneously consuming with approximately 5 kg of Flaxseed oil 7 kg of protein and 31 kg carbohydrates nearly as intermixed as in the fruits of the Primate diet and as much fiber as feasible. Obviously the equivalent ~ (45+28+124) thousand kcal will have to be spread over ~100 days for a total of ~2000 kcal/day and boil down to about 50 g flaxseed oil, 70 g protein and 300 g carbohydrates per day. In connection with this program the following is of interest: For centuries, empirical folk healing arts in Germany have been combating apparent ALA deficiency symptoms along similar lines with considerable success with flaxseed oil – cottage cheese mixture "wonder cures" (16). Beginning with Justus von Liebig leading biochemists later demonstrated the synergy of seed oils and proteins in the human body. More recently, since the 1950ties Johanna Budwig, pharmacologist and practitioner of natural healing arts has published early research on polyunsaturated oils (17) and an oil-protein diet cook book (15), see also reference (18). In the maintenance diet, after the initial high alpha-linolenic pulse, one may want to adjust the fatty acid profile to 30% palmitic, 23% linoleic, 16% alpha linolenic, 15% oleic (and 16% others) and continue to *attempt to loosely approximate* the model ancestral fat, protein, carbohydrate, digested and undigested fiber 5 : 7 : 14 : 17 : 17 weight ratios in wild fruit. Unfortunately, in cultivated fruit, for example a mango, they are very different, 0.45 : 0.45 : 12 : 1.7 : ~0 grams in 100 g of fruit pulp, and would have to be supplemented. Raw vegetables, nuts, eggs, a little wild deer meat, baked potatoes (19) and unprocessed oils may do the job.

Nancy Lou Conklin and Kevin Hunt reviewed the manuscript and offered valuable comments. Katharine Milton kindly reviewed parts of the MS. My wife Diana read the MS and made valuable suggestions. The work was supported in part by Boeing funds.

[10] Topping, D. L. & Clifton, M. P. (2001) Short-Chain Fatty Acids and HumanColonic Function *Physiological Reviews* **81**, 1031-1064

[11] Milton, K & Dintzis, F. (1981) *Biotropica* **13,** 177-181

[12] H. Ti Tien and Angelica L. Ottova (2001) The lipid bilayer concept and its experimental realization: from soap bubbles, kitchen sink, to bilayer lipid membranes *Journal of Membrane Science* **189**, 83-117

[13] Javier S. Perona et al. (2000) Influence of different dietary fats on triacylglycerol deposition in rat adipose tissue *British Journal of Nutrition* **84**, 765-774

[14] Ashima K Kant (2000) Consumption of energy-dense, nutrient-poor foods by adult Americans: nutritional and health implications. *American Journal of Clinical Nutrition* **72**, 929-936

[15] Budwig, J. (1994) *The oil-protein diet cookbook* (Apple Publishing Co., Vancouver, BC)

[16] Hell, C. (2000) *Natur & Heilen* May pp. 18-24

[17] Biology of fats. V. Paper chromatography of blood lipids, the cancer problem and fat research. Kaufmann, H. P. & Budwig, J. (1952) Fette u. Seifen **54,** 156-65. (Translation available from CAS)

[18] A. P. Simopoulos and J. Robinson (1999) *The Omega Diet* (HarperPerennial)

[19] Hatley T, and J Kappelman (1980): Bears, pigs, and plio-pleistocene hominids: a case for the exploitation of belowground food resources *Human Ecol.* **8,** 371-387
7